\documentstyle[12pt,aaspp4,epsf,psfig]{article}

\slugcomment{Draft \today}

\begin{document}

\title{The Mass Distribution of Stellar Black Holes}

\author{ Charles D. Bailyn\altaffilmark{1}}
\affil{Yale University, Department of Astronomy, 
              P.O. Box 208101, New Haven, CT 06510-8101\\
E-mail: bailyn@astro.yale.edu}

\author{Raj K. Jain \& Paolo Coppi}
\affil{Yale University, Department of Physics, 
              P.O. Box 208101, New Haven, CT 06510-8101\\
E-mail: raj.jain@yale.edu, coppi@astro.yale.edu}

\author{Jerome A. Orosz}
\affil{Pennsylvania St. University, Dept. of Astronomy \& Astrophysics,
525 Davey Lab, University Park, PA 16802-6305 
              \\
E-mail: orosz@astro.psu.edu}

\altaffiltext{1}{National Young Investigator}

\begin{abstract}

We examine the distribution of masses of black holes in 
transient low mass X-ray
binary systems.  A Bayesian analysis 
suggests that
it is probable that six of the seven systems with measured mass functions
have black hole masses clustered near seven solar masses.  There appears
to be a significant gap between the masses of these systems and those of
the observed neutron stars.  The remaining source, V404~Cyg, has a mass
significantly larger than the others, and our analysis suggests that it is
probably drawn from a different distribution.  Selection effects do not
appear to play a role in producing the observed mass distribution, which
may be explained by currently unknown details of the supernova explosions
and of binary evolution prior to the supernova.

\end{abstract}

\keywords{binaries: spectroscopic --- 
black hole physics --- supernovae --- X-rays: binary}

\section{Introduction}

The strongest case for the existence of black holes in nature is provided
by a subclass of X-ray transients.  The strong episodic X-ray emission
from these binary stars demonstrates the existence of an accreting compact
object.  Radial velocity measurements of the companion star can be used
to determine the mass function, which is a strict lower limit to the
mass of the compact accretor.  In a number of cases, this lower limit
is above the upper limit of neutron star stability of $3M_{\odot }$
(McClintock \& Remillard 1986, Casares \& Charles 1992,
Remillard, McClintock \& Bailyn 1992, Bailyn et al.
1995, 
Filippenko
et al. 1995,
Remillard et al. 1996). 
In these cases the compact object must be a black hole.

Studies of the supernovae explosions which presumably give rise to these
black holes have now progressed to the point where meaningful statements
can begin to be made about the expected mass distribution of the black holes.  
Studies of galactic chemical evolution strongly imply that stars with 
initial masses above $\approx 30M_{\odot }$ must ``swallow'' many of their
heavy elements during (or shortly after) the supernova event, and therefore
should form relatively massive black holes
(Maeder 1992, Timmes, Woosley \& Weaver 1995).  Detailed models of the 
supernovae explosions themselves (e.g. Timmes, Woosley \& Weaver 1996) can
reproduce this result by applying a fixed amount of kinetic energy to the
outer layers of the star, and determining how much material recollapses onto
the central object.  However, there remain many undertainties in the 
relation between the initial mass of the star and the mass of the compact
remnant left behind by the supernova.  The amount of kinetic energy 
available, how it is transferred to the ejected material,  details of the
pre-supernova evolution of massive stars (especially relating to convection
and mass loss) and the possible influence of a binary companion are all
poorly understood.  In this paper, therefore, we take an empirical approach
--- we attempt to use the available observational evidence for stellar 
black holes with low mass companions (high mass systems such as Cygnus~X-1
presumably having followed a different evolutionary path), and see what 
constraints can be made upon the underlying distribution of black hole
masses.

The mass of the black hole primary  $m_1$ 
is related to the observed mass function $f(m)$ in
the following way:
\begin{equation}
m_1={{f(m)(1+q)^2}\over {\sin ^3 i}}
\end{equation}
In order to determine the mass of the black hole, one needs to know
the mass ratio $q$  and the inclination 
$i$ of the binary system.  These parameters
can be measured, although somewhat more indirectly than the mass function,
as discussed in section~2 below.  Section 2 also presents a
compilation of the data obtained on those 
low mass X-ray transients with measured
mass functions which do not display X-ray bursts (generally considered
to be a neutron star signature).  In section 3, we use these data to
examine the distibution of black hole masses following the Bayesian
analysis employed by Finn (1994) in his study of neutron star masses.
Our results suggest that V404~Cyg is not drawn from the same population
as the other six sources, and that the group of six have black
hole masses which cluster around $7M_{\odot }$ with a significant gap 
between their distribution and that of the neutron stars.  
In section 4 we speculate about 
the implications of our results.

\section{Constraints on Black Hole Masses}

As can be seen in Equation~1, the mass ratio and inclination of the binary 
system are required in addition to the mass function to determine the mass
of the black hole primary.  These quantities have been measured in a 
variety of ways, which are discussed in sections 2.1 and 2.2.  Section
2.3 evaluates the currently available data on each of the seven systems
being considered.

\subsection{Determining the Mass Ratio}

The most conceptually straightforward way to determine
the mass ratio of a binary system is by
measuring a velocity curve for both components of the binary system.
Therefore several attempts have been made to construct an orbital
velocity curve for the emission lines
emanating from the accretion
disks in black hole binaries (e.g. Orosz et al. 1994). 
Such a velocity curve should track the motion of the primary; the ratio of
its amplitude to that of the velocity curve of the absorption spectrum
of the secondary determines the mass ratio.
However, the emission line
velocity curve is observed to
lag behind its expected phase, and thus cannot
reflect the true motion of the primary (Orosz et al. 1994).  Thus this
method, while conceptually simple, is suspect in practice.  Curiously,
despite the phase lag,
the results obtained by naively applying this method agree with
those of other methods.

A more reliable method for determining the mass ratio is to observe the
rotational broadening of the lines from the secondary star.  
The secondary
fills its Roche lobe, and the ratio of its effective radius to the 
orbital semi-major axis is therefore determined by the mass ratio.
By comparing the projected orbital velocity of the secondary (as determined by
the velocity curve) and the 
projected rotational velocity observed from the line
broadening, the mass ratio can be determined.  This procedure is thought
to be reliable, but the observations require relatively high dispersion
and signal to noise,
and are thus not feasible for the faintest systems.

The secondaries of these system are generally undermassive for their
spectral types when compared to typical main sequence stars.  Presumably
this is because they are not in thermal equilibrium, due to the ongoing
mass loss.  In cases where the rotational broadening cannot be observed,
crude limits on the mass of the secondary can be set by assuming that
the secondary is somewhat less massive than a main sequence star of
similar spectral type.

\subsection{Determining the Inclination}

The orbital inclination can in principle be determined by observing
the ellipsoidal variability of the Roche lobe filling secondary.  This
has been done with high precision in the case of GRO~J1655-40 (Orosz \& Bailyn
1997).  In this case, the shapes of the lightcurves
were sufficiently accurately determined that the mass ratio could also
be determined from the models.
However in other cases, the lightcurves are
modified by several effects, which can only be modelled approximately.

There is a contribution of light from the disk, which is approximately
constant throughout the orbit, and thus decreases the ellipsoidal
amplitude for a given inclination.  The disk fraction can be determined
by comparing the depths of the absorption features (produced by the
secondary) with those of single stars, but the disk can vary from 
observation to observation, and also as a function of phase, so the
correct disk fraction to apply is not in general well-determined.
Another source of distortion for the lightcurves may be star spots on
the rapidly rotating secondaries, which may be particularly important
for late spectral types.

It is often assumed that light curves in the IR will be free of disk
contamination, so ellipsoidal modelling of IR lightcurves is a favored
method of determining the inclination.  However it should be noted that
in GRO~J1655-40, the disk is actually redder than the secondary star in 
quiescence, so the assumption that the IR is relatively free of disk
contamination may not be reliable.

\subsection{Specific Systems}

Here we briefly describe the data available on each of the seven systems.
The results are summarized in Table~1, and in Figure~1, which shows
the mass limits derived by considering the extremes of the errors and
ranges listed in Table 1.  The systems are listed in order of decreasing
mass function.

\bf GS~2023+34 = Nova Cyg 1938/1989 = V404~Cyg: \rm  This source
has the highest measured mass function of any compact object (Casares \&
Charles 1994) and
is therefore the strongest case for a black hole.  Casares \& Charles 
(1994) also derive the mass ratio from rotational broadening, and Pavlenko
et al. (1996) confirm the inclination limits found by
Shahbaz et al. (1994) 
from ellipsoidal modelling of the IR light
curve.

\bf GS~2000+25 = Nova Vul 1988: \rm The best values for the mass function
and the mass ratio from rotational broadening were reported by Harlaftis
et al. (1996).  The limits on the inclination are derived from ellipsoidal
modelling of the IR light curve by Beekman et al. (1996).

\bf H~1705-25 = Nova Oph 1977: \rm The best value for the mass function
of this system is given by
Filippenko et al. (1997).  
The mass ratio is adopted to fit the limits on rotational broadening 
given by Harlaftis et al. (1997).
The range of inclinations is
from ellipsoidal modelling of the optical light curve by Remillard
et al. (1996).

\bf GRO~J1655-40 = Nova Sco 1994: \rm This source displays remarkably
precise ellipsoidal variations in quiescence, which have allowed us
to determine the mass ratio and inclination of the system to unprecedented
precision (Orosz \& Bailyn 1997).  The excellent fits are presumably due
to the relatively early spectral type of the secondary (F5) which decreases
the effects of starspots, and the very small contribution from the
disk ($\le 5\% $).  This is the only case among these sources in which
the shape of the ellipsoidal lightcurve is sufficiently well-determined
to provide a strong constraint on the mass ratio.  The formal error
on the inclination quoted by Orosz \& Bailyn is only $\pm 0.1^o$, but we
use a larger range here which encompasses several other local minima
in $\chi ^2$, even though these minima are formally several sigma less
significant than the overall best fit at $i=69.50^o \pm 0.08$.

\bf GS~1124-68 = Nova Mus 1991: \rm The most recent results for 
the mass function and inclination of this
source are given by Orosz et al. (1996).  The mass ratio is determined from
the observed rotational velocity by Casares et al. (1997) and agrees with
the results from the emission line velocity curve given by Orosz et al. 
(1994).

\bf A~0620-00 = Nova Mon 1975 = V616~Mon: \rm  This source was the first of its
class for which a mass function was
determined (McClintock
\& Remillard 1986).  The source has been extensively studied over the
past decade by many authors.  However there remain sharp disagreements
over the inclination.  Ellipsoidal variability measurements in the IR
give inclinations in the range of $31^o\le i \le 54^o$ (Shahbaz et al.
1994a), while
lightcurves in the optical and rotational velocity measurements yield
much higher inclinations (e.g. Haswell et al. 1993).  
Rather than attempt to resolve this
discrepancy here, we simply adopt the entire range of reported results.
This large range of inclinations yields a correspondingly large range
in values for the primary mass.  In contrast, all of the various 
measurements of the mass ratio yield consistent results.

\bf GRO~J0422+32 = Nova Per 1992: \rm This is the only source included
here with a mass function significantly less than $3M_{\odot }$
(Filippenko et al. 1995).  However
it displayed no X-ray bursts, and there are a number of indications that
the inclination is relatively low, and that therefore the primary mass
is in the black hole range.  However, attempts to determine the inclination
by studying ellipsoidal variations
have been ambiguous, probably because the effects of starspots
are particularly acute for this relatively late type secondary (M0).
Orosz \& Bailyn (1995) report an inclination of $\approx 45^o$, but
this conclusion was undermined by subsequent data which revealed a
smaller ellipsoidal amplitude when the source was $0.2$ magnitudes
fainter, which goes in the wrong direction for a change in disk
contamination (Callanan et al. 1996).  The results of Callanan et al. (1996)
clearly indicate that $i<45^o$; as a lower limit, we adopt
an inclination which
provides just enough ellipsoidal variability to produce the 
smallest observed ellipsoidal modulation in the complete absence of
disk contamination.

\section{Application of Bayesian Statistics}

Figure 1 shows the ranges over which the primary masses of the seven
sources can vary, given the observational constraints.  It is important
to note that these mass limits cannot be interpreted as coming 
from gaussian distributions that are characterized by some ``sigma.''  The 
heterogeneous nature of the observed quantities, and the non-linear ways
they enter into the calculation of the primary mass, mean that the
probability for a black hole to be at a given point between the
plotted limits varies in a strongly non-gaussian manner. Therefore,
to learn more about the black hole masses and, in particular, to 
see what parent mass distribution they could have been drawn from,
we have performed a Bayesian analysis  similar to that used by 
Finn (1994) to study the distribution of neutron star masses
in binary systems.  

Bayesian analysis is a powerful tool
that allows one to quantify the consistency of a particular model,
in this case a parent distribution of black hole masses, 
with a set of data taking into account 
any ``prior" knowledge (biases) we might have concerning the systems being
studied along with the observational uncertainties involved
in obtaining the data.   In brief (e.g., see Loredo 1990),
the Bayesian approach does not assign error bars to 
data but rather starts with a model that is supposed to 
describe the system that produced the data. It then asks what is 
the likelihood for that particular model to be relevant or 
correct in the first place (based on any prior knowledge or biases
we may already have concerning the system),
and finally it asks what is the likelihood 
the data points could have been generated by the 
model given what we understand about the measurement process 
that produced the data points.
To those new to Bayesian analysis, this approach often seems backwards 
and troubling since one usually thinks of the data as the starting point 
for any analysis and also feels strongly that the 
final answer should be objective, i.e., not depend
on the prior knowledge of the observer.
However, the process of Bayesian inference actually matches closely
the process of how the scientific community arrives
at what it considers to be scientific knowledge
(e.g., an apparently significant experimental result is ignored 
because everyone knows the detector ``is always broken")
and is ``honest" in the sense that all the 
underlying assumptions are spelled out explicitly. 
Also, Bayesian analysis is not quite as subjective 
as it might first appear.  For many classes of problems, one can 
show that there exists a unique set of prior knowledge or assumptions
that is ``reasonable'' in the sense that it is  ``minimally biased''
or contains the ``least amount" of prior information.
Although the approach of starting with 
a well-specified  model instead of the data 
can be computationally cumbersome, it makes maximum use of the
available information and often allows one to make 
significant, quantitative statements about model-data consistency 
even when the data are sparse, as is the case here.
But it should be remembered that 
conclusions based on a Bayesian
analysis always depend critically on the assumed models and 
prior knowledge (biases).  

\subsection{Method}

To start our Bayesian analysis, we must put down the assumptions
and models for our particular problem. Since we have only seven black
hole systems for study, we do not expect to be able to make very detailed
statements about any particular parent distribution for black hole masses. 
Accordingly, we shall only consider a very simple model where the masses
of the black hole primaries are uniformly distributed 
between between some lower and upper mass limits, $m_l$ and $m_u$ 
respectively. Our set of prior knowledge, which we denote
${\cal I}$, includes the available information (discussed above) for the
inclinations and mass ratios of individual systems, the assumption 
that our model mass distribution is correct, broad bounds on the 
possible values of $m_l$ and $m_u,$ and the assumption that all values 
of $m_l$ and $m_u$ are {\it a priori} equally probable within those bounds,
provided $m_l \le m_u.$ Given this prior knowledge, our goal is then 
to determine the probability density function $P(m_l,m_u|f_n,{\cal I}),$ 
which tells us how likely it is for the
mass distribution specified by a particular choice of
of $m_l$ and $m_u$ to be the correct one given
the set of observed mass functions ${f_n}$ (our data). We will express
our results in terms of probability contours in the $(m_l,m_u)$ plane.

Bayes' Theorem, applied to this problem, states that
\begin{equation}
P(m_l,m_u | \{ f_n\} , {\cal I}) = 
{{P(m_l,m_u | {\cal I}) P(\{ f_n\} | m_l,m_u, {\cal I})} \over
{P(\{ f_n \} | {\cal I})}} .
\end{equation}
The quantities on the right hand side of equation 2 are interpreted in
the following manner.  The first term of the numerator
$P(m_l,m_u | {\cal I})$
is the probability that a given combination of $m_l$ and $m_u$ is allowed 
by our prior information.  We will take this probability to be uniform
for $m_l \le m_u$ and $0.5M_{\odot } \le m_l,m_u \le 30M_{\odot }$,
and zero otherwise.  This mass range was chosen to encompass the entire
reasonable range of black hole masses in transient systems.
Since $P$ must integrate to unity over all $m_l$ and $m_u$, we find that
\begin{equation}
P(m_l,m_u | {\cal I}) = {2\over {(30-0.5)^2}} 
\end{equation}
within the bounds given above, where we have chosen to measure the
$(m_l,m_u)$ plane in units of solar masses.

The second term in the numerator, 
$P(\{ f_n\} | m_l,m_u, {\cal I})$, is the probability of obtaining a
specific set of observations $\{ f_n \}$ given our prior assumptions,
and specific values of $m_l$ and $m_u$. Since the observations of the
individual black holes are independent of each other we can factorize
$P(\{ f_n\} | m_l,m_u, {\cal I})$ as follows:
\begin{equation}
P(\{ f_n\} | m_l,m_u, {\cal I}) = \prod _i P(f_i | m_l,m_u, {\cal I})
\end{equation}
where $f_{i=1,n}$ are the mass functions obtained for the individual 
sources.
The product rule of probability yields
\begin{equation}
P(f_i | m_l, m_u, {\cal I}) = \int d\hat f_i P(f_i | \hat f_i,{\cal I})
P(\hat f_i | m_l,m_u,{\cal I})
\end{equation}
where $\hat f_i$ is the true value of the mass function, and $f_i$ is the 
observed value.  We will assume that the relationship between the true
and observed values of the mass function is given by a gaussian 
probability distribution with $\sigma $ equal to the error in $f_i$
quoted by the observers.
Thus
\begin{equation}
P(f_i | \hat f_i, {\cal I}) 
= {\exp [- {1\over 2} ({f_i-\hat f_i) \over {\sigma }})^2]}/2\pi \sigma .
\end{equation}

The function 
$P(\hat f_i | m_l, m_u, {\cal I})$ is the likelihood for the true value
of the mass function to be $f_i$  given  ${\cal I}$ and 
particular values of
$m_l$ and $m_u$.  Since the black hole mass 
\begin{equation}
m_i=\hat f_i (1+q)^2/\sin ^3i,
\end{equation}
we can change variables and write
\begin{equation}
P(\hat f_i | m_l,m_u, {\cal I}) = \int \int di dq P(m_i)P(i)P(q)
{{(1+q)^2} \over {\sin ^3 i}}
\end{equation}
where $P(m_i)$ is uniform for $m_l\le m_i \le m_u$ and zero otherwise,
$P(q)$ is a gaussian given by the mean value and error quoted in Table~1,
and $P(i)\propto \sin i$ (i.e. $P(\cos i)$ is uniform) between the bounds
given in Table~1 and zero elsewhere, normalized to yield unity when 
integrated over all angles.  If we compute $m_i$ for each value of
$q$ and $i$ using equation~7, the integral
in equation~8 can be computed using standard methods for any given value
of $\hat f_i$, and these results can be used to compute 
$P(\{ f_n\} | m_l,m_u, {\cal I})$ using 
equations~6 and~4.

Finally, the remaining term in the right hand side of equation 2,
$P(\{f_n\}|{\cal I}),$  is  the ``prior predictive probability'' or 
``global likelihood'' 
of the model under consideration.  This term represents the
 probability that the observed 
set of $\{f_n\}$ can be obtained given 
our set of prior knowledge and assumptions, ${\cal I}$. For our problem,
this represents the probability that the black hole mass distribution
for our seven systems can be correctly modeled as a uniform
distribution between two mass limits, $m_l$ and $m_u$, constrained
to lie between $0.5M_{\odot }$ and $30M_{\odot }.$
In most Bayesian analyses, this term is to be
regarded as a normalization constant. The left hand side of equation 1 must 
by definition have a value of unity when it is integrated over all
possible values of $m_l$ and $m_u$, and this
requirement fixes $P(\{ f_n \} | {\cal I})$ if the other quantities 
on the right-hand side of equation 2 are known, ie., 
\begin{equation}
P(\{ f_n \}| {\cal I}) = \int_{0.5}^{30} \int_{0.5}^{30} d m_u d m_l
 P(\{f_n\} | m_l,m_u,{\cal I}) P(m_l,m_u | {\cal I}).
\end{equation}

The global likelihood of a model is a very useful quantity as it allows
one to objectively compare classes of models. Larger values of the global
likelihood imply that the underlying model is more likely to be
correct.  Specifically, the ratio of the global likelihood of two models
can be shown to constitute an ``odds ratio'' which compares one model
to the other (Loredo 1990).  To see this result,
let us
assume that we add an extra piece of information, ${\cal I^\prime}$ to 
our prior knowledge.  This new information states 
that we assume that our data could be
explained by some number $N$ of possible models and that only one
of these models, we don't know which,
is actually the right one. Then using Bayes theorem, we can write
\begin{equation}
P(M_i|\{f_n\},{\cal I}, { \cal I^\prime})=
P(M_i|{\cal I}, {\cal I^\prime}){P(\{f_n\} | M_i, {\cal I}, 
{\cal I^\prime}) \over
P(\{f_n\} | {\cal I}, {\cal I^\prime})}
\end{equation}
where $M_i$ refers to the $i^{th}$ of the models we are considering
and $P(M_i| {\cal I}, {\cal I^\prime})$ is our prior probability 
for that model
to be correct. 
The expression $P(\{f_n\} | M_i, {\cal I}, {\cal I^\prime})$ is 
the probability
that we reproduce the data given (i) our prior set of knowledge
${\cal I}$; (ii)  that a  set of possible models ${\cal I^\prime}$
exists; and (iii) that of these models, $M_i$ is the correct one.
Since we assumed that only one model could be correct, clearly this
probability is independent of the existence of any alternative models,
$M_{i\ne j}$. Hence, this probability is the same as
 $P(\{f_n\} | M_i, {\cal I}),$ 
the probability that we can reproduce our data using model $M_i$ 
constrained by our original set of prior knowledge, i.e., it is  
the global likelihood for model $M_i$ (calculated in the manner described 
above by integrating over all the possible parameters of the model).
Knowing this we can now calculate the odd's ratio, $O_{ij},$ favoring  one
of the models, $M_i,$ over another, $M_j,$
\begin{equation}
O_{ij} = {P(M_i | {\cal I^\prime}) \over P(M_j | {\cal I^\prime}}
 { P(\{f_n\} | M_{i}, I) \over P(\{f_n\} | M_{j},I)}.
\end{equation}
In other words, the probability favoring a model as a whole is 
proportional to its prior probability times its global likelihood.
In what follows, we shall assume that we have no prior knowledge
of which $M_i$ is correct and set all the $P(M_i | {\cal I^\prime})$
equal. Thus the odds favoring one model over another is 
simply the ratio of the global likelihoods.

Our overall procedure used to compute all the required probabilities
is as follows.
We take a grid of values of 
$m_l$ and $m_u$ in the range $0.5M_{\odot } \le m_l \le m_u \le 
30M_{\odot }$.  For each pair of $(m_l,m_u)$ values, we compute
$P(f_i | m_l, m_u, {\cal I})$ for each source
by integration as described above.
The results for each source are then multiplied together to get
 $P(\{ f_n|\} | m_l,m_u,{\cal I})$ for each grid point in $(m_l,m_u)$ space. 
We next sum
these values over $m_l$ and $m_u$, and determine a value for the 
global likelihood 
$P(\{ f_n\} | {\cal I})$ such that the sum of 
$P(m_l,m_u | \{ f_n\} , {\cal I})$ is unity, as required.  

\subsection{Results}

The results of applying this method to the full data set of seven objects
is shown in Figure~2.  The contour enclosing 95\% of the probability
covers a region with $11\le m_u/M_{\odot } \le 17$, and 
$m_l/M_{\odot } \le 7$.  The maximum likelihood point is at
$m_l=5.5M_{\odot }$ and $m_u=11.7M_{\odot }$.

A more intriguing result is obtained by modifying one of the assumptions
implicit in ${\cal I}$. Instead of assuming all the black hole masses
are drawn from the same parent mass distribution, we assume that
the masses of only six of the objects are drawn from this distribution 
and that the mass of the seventh comes from a different one. For simplicity,
we will assume that the mass of the seventh is in fact known to have some
value consistent with the data so that the probability $P(f_7 | {\cal I})$
equals unity and that object effectively drops out of the computation.
Assuming that each system in turn has a known mass, we compute a
corresponding global likelihood for the resulting model. (This is 
the same likelihood as before except we now consider only the 
six remaining objects).  The surprising set of likehood values we 
obtained is shown in Table~2.
By comparing these global likelihood values to 
each other and to the one we obtained above, 
we can ascertain
whether a particular model is ``favored'' relative to the others.
In most cases, removing one object made little difference to
the either the global likelihood or the shape of the contours
for the model probability distribution (as a function of $m_l$
and $m_u$). However, when GS~2023+34 (V404~Cyg) was removed, the change
was dramatic (see Figure~2).  

When sources other than V404~Cyg were left out,
the model global likelihoods rose by only factors of a few.
(They rose presumably because our
assumption that $m_u$ could be as high as 30$M_{\odot }$ was wrong ---
with each additional source this assumption becomes less plausible).
By constrast,
when V404~Cyg was removed, the global likelihood of the model
 rose by over two orders
of magnitude.  This strongly suggests that 
V404~Cyg is indeed drawn from a different distribution than the other sources.
It should be noted that V404~Cyg is also unusual in having a longer
orbit and more evolved secondary than the other sources.

Looking at the probability contours when V404~Cyg is left out,
the
maximum likelihood values of $m_l$ and $m_u$ become much closer 
($m_l=6.91$ and $m_u=6.97$).  Indeed, there is considerable probability
that $m_l=m_u$, i.e. that the distribution can be plausibly modelled
by a single black hole mass near $7M_{\odot }$.  Figure~3 shows the 
probability that $m_l$ and $m_u$ are within some value of each other.
This plot was generated by integrating over the normalized probability
distribution for all $m_u-m_l \le \Delta m$.  When all seven sources are
considered, the distribution cannot be narrower than $4M_{\odot }$ wide
(i.e. $\Delta m \ge 4M_{\odot }$).  When V404~Cyg is not included,
over 50\% of the probability has $\Delta m \le 2M_{\odot }$.  This result
suggests that the possibility that many black holes have masses in a small
range near $7M_{\odot }$ should be seriously considered.

Another effect of leaving out V404~Cyg is that the likely values for $m_l$ 
becomes  much higher.  
Figure~5 shows the probability that $m_l$ is below some value.
As can seen, there is a 95\% probability that $m_l>3M_{\odot }$, and
a 90\% probability that $m_l>4M_{\odot }$.  Given the upper limit of
$\le 1.7M_{\odot }$ for the observed neutron star distribution, this
result implies a considerable gap in the observed mass distribution of
compact objects.

\section{Discussion}

The mass distribution of stellar mass black holes can in principle be
predicted by models of the evolution and supernova explosions of massive
stars.  Timmes et al. (1996 --- hereafter TWW) for example find that 
compact remnants of supernovae become significantly more massive than the 
pre-supernova iron core when the progenitor's initial mass exceeds
$30M_{\odot }$.  Above this mass, the kinetic energy of the explosion is
insufficient to unbind the entire mantle and envelope of the star, so
some of the outer regions fall back hole the core, increasing the mass of
the remnant.  Studies of galactic chemistry confirm that stars above 
$30M_{\odot }$ cannot return their entire outer regions to the interstellar
medium (Maeder 1992, Timmes et al. 1995).  
Taking the reasonable assumption that the applied kinetic
energy is only weakly dependent on the progenitor's initial mass, TWW find
a smooth
monotonic relation between remnant and progenitor mass (see their Figure
1).  If the initial mass function is weighted toward lower mass stars, as
is almost certainly the case, the resulting mass distribution of black 
holes
should be smooth, and strongly biased toward masses near those of the 
neutron stars.

While a sample including only seven examples cannot provide unambiguous 
statistical results, our work strongly suggests that the expected 
distribution is not observed.  As discussed above, it is likely that one
of the seven sources is not drawn from the same distribution as the other
six.  These six black holes may well be tightly clustered in mass near
$7M_{\odot }$, althought broader distributions are also possible.  Finally,
there is a significant gap between the lower mass limit of this 
distribution and the upper limit of the observed neutron star masses
given by
Finn (1994).  The expected pile-up of sources at masses near to that of the 
neutron stars appears to be ruled out.  Thus, although the small number of
sources prevents us from drawing completely compelling conclusions, the
evidence against the kinds of distributions implied by the work of TWW
is strong enough to warrant serious consideration of the ways in which
those distributions could be significantly modified.

A possible caveat is that this last conclusion might arise from 
observational selection effects. At present, however, we cannot
think of any that would prevent systems with black hole masses
near the neutron star upper limit from being detected.
These sources are easy to identify --- in outburst they are among
the brightest objects in the X-ray sky.
Also, the clear observation of similar systems containing accreting 
neutron stars strongly suggests that the mass of the compact object is
not critical in the identification of these sources.  Note that we have 
included \it all \rm transient systems with measured mass functions in
our sample, except those like Cen~X-4 (McClintock \& Remillard 1990) 
which display type I X-ray
bursts, which are convincing signatures of the presence of a neutron
star.  In particular, we included GRO~J0422+32, which is often left off
lists of confirmed black holes because its mass function is well below
the limit of $3M_{\odot }$ generally taken to be the firm upper limit 
for neutron star stability.  It is conceivable that some effect
suppresses the disk instability cycle over some mass range of the
primary star, thus preventing the transient behavior needed to both
identify the source in the first place, and then subsequently measure the
mass function.  Once again, however, the existence of transient sources
with neutron star primaries and recurrence timescales similar to 
or smaller than those of the black hole systems makes this solution 
implausible.

The physical effects which might influence the black hole distribution can
be divided into two classes: those which involve the supernova explosion
itself, and those related to the binary nature of the observed systems.
Regarding the supernova explosions, if  the relation
between the amount of fall-back material to the mass of the precursor were
close to a step function, then one might imagine that if \it any \rm
significant amount of material were to fall back, there would be enough
to bring the total remnant mass up to $\approx 7M_{\odot }$.
In this
case, even if the amount of fallback material increased with increasing
precursor mass after the initial step, one would still expect a strongly
peaked distribution due to the steep expected mass function of massive
stars.  One conceivable scenario which might lead to such a step function
is if some chemical transition within the precursor immediately prior to
the supernova occurred just exterior to $\approx 7M_{\odot }$.  Then the
higher density and atomic number interior to the transition might result in
precisely that portion of the star recollapsing, while the outer regions
are expelled.  Unfortunately, evolutionary calculations of very massive
stars are subject to a number of important uncertainties (Woosley
\& Weaver 1995) 
particularly relating to mass loss rates, so this suggestion is hard to
evaluate.

It may be that the observed distribution of black hole masses is not a
result of type II supernova explosions in general, but rather a consequence
of the binary nature of the observed systems.  All of the orbital periods
are sufficiently small that considerable mass transfer must have occurred 
prior to the supernova explosion.  Since the precursor of the supernova
was almost certainly more massive than its companion, it would have filled
its Roche lobe first, resulting in dynamically unstable mass transfer.  
This in turn would result in a common envelope configuration, leading to
the expulsion of much of the outer envelope of the more massive star and
a dramatic decrease in the orbital separation on a very short timescale.
Such abrupt mass loss has been shown to dramatically change the
nature of the subsequent supernova explosion (Brown, Weingartner \&
Wijers 1996).  However it is not 
clear how this set of circumstances could lead to a narrow range of remnant
masses, or to a gap, since the effects of common envelope evolution are 
strongly
dependent on the initial binary separation, which presumably is broadly
distributed.

The mass of the black hole may also be influenced by mass accretion 
subsequent to the supernova event.  If this effect changed the mass of the
primary significantly, one would expect to see broadly distributed masses
for the black hole, since the system should have undergone varying amounts of 
accretion after the formation of the compact object.  Such a broadened
distribution cannot currently be ruled out, but confirmation of the 
suggested
sharp mass distribution  would argue strongly against significant post-SN
mass enhancement.  In most cases there are also evolutionary arguments 
against post-SN mass enhancement.  Five of the seven systems in our sample
contain late-type main sequence secondaries.  These stars have masses
$<1M_{\odot }$, and it would require considerable fine tuning if they had
all begun with much more massive secondaries.  GRO~J1655-40 has a secondary
in the Hertzsprung gap --- considerable ingenuity is required to reconcile
this with the required low mass transfer rate for transient behavior
(Kolb et al. 1997), and it is not clear how much mass transfer might have
already taken place.  The remaining system is V404~Cyg, whose secondary
is at the base of the giant branch.  In this case, considerable material
may indeed have been transferred, which might conceivably account for the 
anomalously high mass of this particular black hole.

Thus, while there are a number of poorly understood effects which might 
alter the distribution of post-supernova remnant masses, it is not 
immediately obvious how these effects could combine to produce the kind of
distribution favored by our analysis.  Obviously, more examples of
black hole systems and better measurements of known systems will result
in more precise observational constraints using techniques
such as the ones outlined here.  But already it appears 
likely that we will have to consider new 
underlying mechanisms for the origin of the Galactic black hole mass
distribution.

\begin{acknowledgements}

We are grateful for conversations with Richard Larson.  CDB and RKJ
acknowledge support from the National Science Foundation (National
Young Investigators program).
\end{acknowledgements}

\clearpage

\begin{deluxetable}{llll}
\tablewidth{0pt}
\tablecaption{Observed Constraints on Black Hole Mass} 
\tablehead{
\colhead{Object} & \colhead{f(m)} & \colhead{$q$} & 
\colhead{$i$}             
}
\startdata
2023+34 & $6.07 \pm 0.05$ & $0.060 \pm 0.005$ & $52\le i \le 60$ \nl
2000+25 & $5.01 \pm 0.12$ & $0.042 \pm 0.012$ & $43\le i \le 74$ \nl
1705-25 & $4.65 \pm 0.21$ & $0.018 \pm 0.016$ & $60\le i \le 80$ \nl
1655-40 & $3.24 \pm 0.09$ & $0.333 \pm 0.010$ & $67\le i \le 71$ \nl
1124-68 & $3.01 \pm 0.15$ & $0.13 \pm 0.04$ & $54\le i \le 65$ \nl
0620-00 & $2.91 \pm 0.08$ & $0.067 \pm 0.010$ & $31\le i \le 70.5$ \nl
0422+32 & $1.21 \pm 0.06$ & $0.049 \pm 0.020$ & $28\le i \le 45$ \nl

\enddata
\end{deluxetable}

\clearpage

\begin{deluxetable}{cccccc}
\tablecaption{Results of Bayesian Analysis}
\tablewidth{0pt}
\tablehead{
\colhead{Sources}   & \colhead{max likely $m_u$} & 
\colhead{max likely $m_l$} & 
\colhead{range $m_{u}$}  & \colhead{range $m_{l}$} & 
\colhead{relative likelihood}  
} 
\startdata

All  & 11.8 & 5.0 & 10.5 $\sim$ 18.0 & 0.5 $\sim$ 6.9 & $1$ \nl

No 0422  & 11.9 & 5.0 & 10.5 $\sim$ 19.4 & 0.5 $\sim$ 6.9 & $2.0$ \nl

No 0620  & 11.9 & 5.1 & 10.5 $\sim$ 18.5 & 0.5 $\sim$ 6.9 & $5.6$ \nl

No 1124  & 12.1 & 5.0 & 10.6 $\sim$ 19.2 & 0.5 $\sim$ 7.1 & $5.3$ \nl

No 1655  & 12.2 & 4.8 & 10.5 $\sim$ 19.5 & 0.5 $\sim$ 6.8 & $4.2$ \nl

No 1705  & 12.1 & 5.2 & 10.5 $\sim$ 19.0 & 0.5 $\sim$ 7.2 & $8.8$ \nl

No 2000  & 11.9 & 4.8 & 10.5 $\sim$ 18.6 & 0.5 $\sim$ 6.9 & $6.3$ \nl

No 2023  & 6.9 & 6.8 & 6.3 $\sim$ 11.0 & 2.6 $\sim$ 7.5 & $134.0$ \nl
 
\enddata

\end{deluxetable}

\clearpage

\begin{figure}
\centerline{
\psfig{file=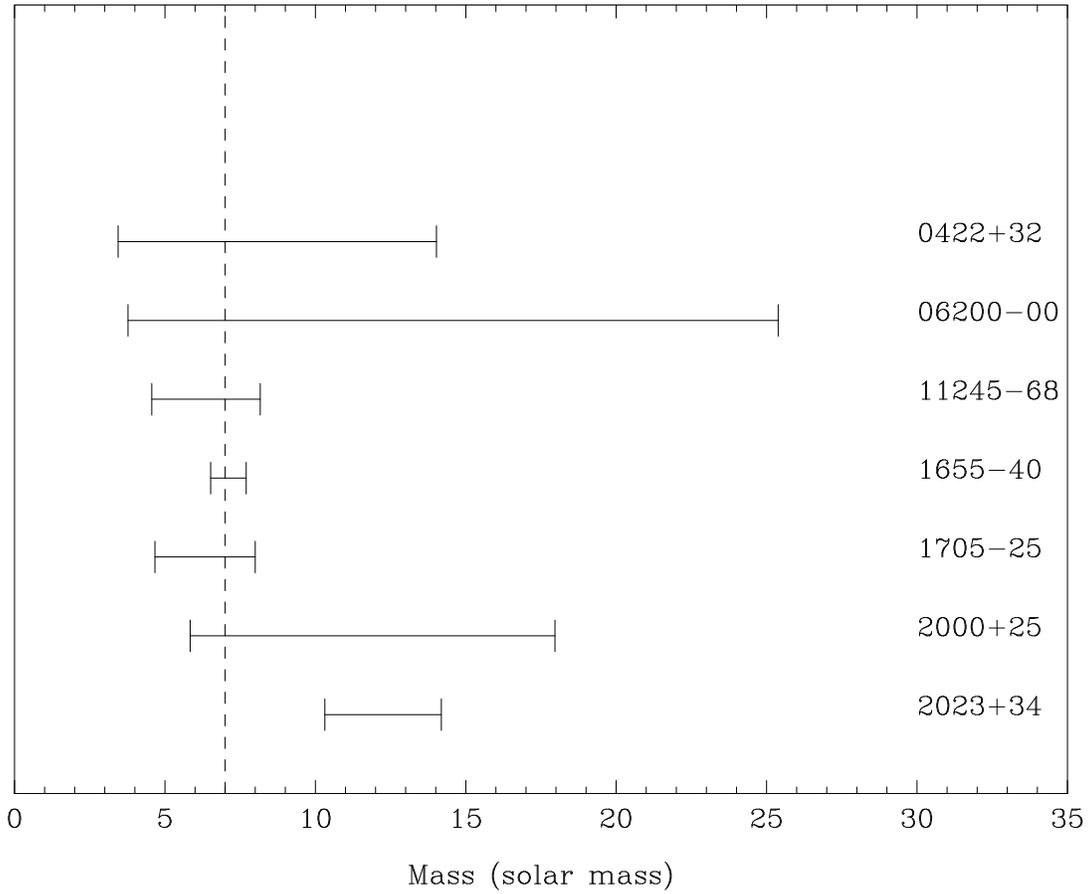,height=15cm,width=20cm,angle=270}
}

\caption[]{Mass ranges for the compact primaries of the seven sources in
our sample.  Note that it is incorrect to think of these ranges as
representing some number of ``sigma'' about a mean value, due to the
strongly non-gaussian nature of the probability distribution.} 
\end{figure}

\begin{figure}
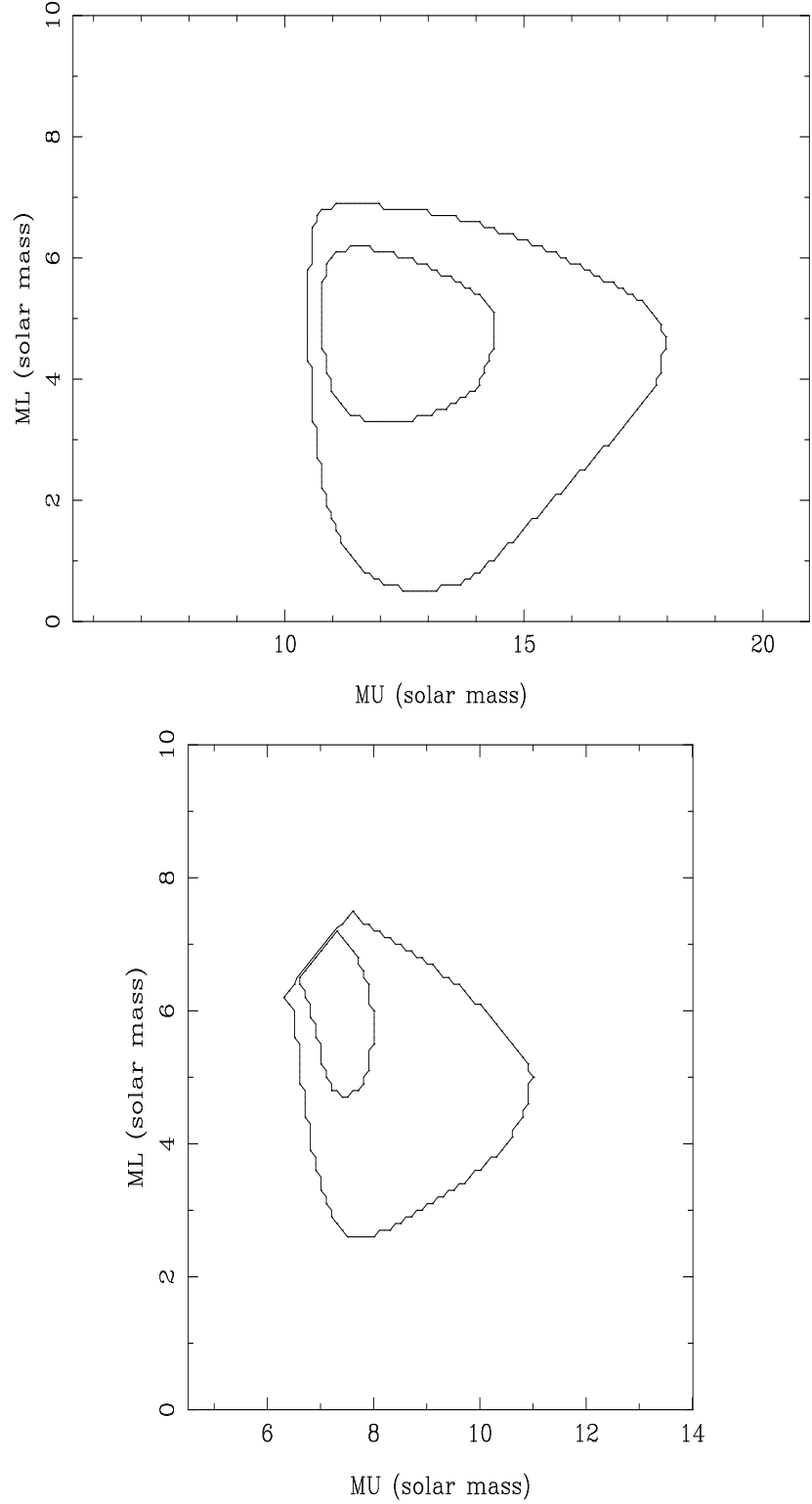


\centerline{
\psfig{file=bc_all_3.ps,height=10cm,width=11cm,angle=270}
}
\centerline{
\psfig{file=bc_no2023_3.ps,height=10.5cm,width=11.5cm,angle=270}
}

\caption[]{Probability contours in the ($m_L,m_U)$ plane.  Inner contour
contains 50\% of the probability; outer contour contains 95\% of the
probability.  Top panel includes GS~2023+34; bottom panel does not.}
\end{figure}

\clearpage

\begin{figure}
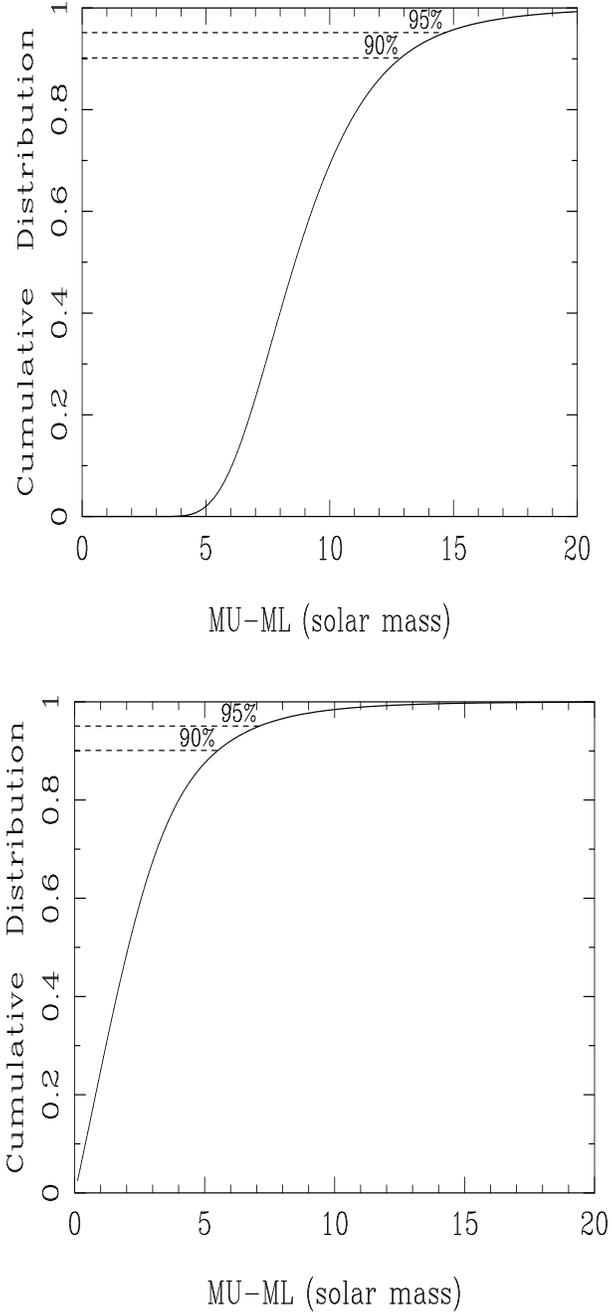

\centerline{
\psfig{figure=bc_all_3_muml.ps,height=9cm,width=8cm,angle=270}
}
\centerline{
\psfig{figure=bc_no2023_3_muml.ps,height=9cm,width=8cm,angle=270}
}

\caption[]{Cumulative probability that $m_U-m_L$ is less than a given
value.  Top panel includes GS~2023+34; bottom panel does not.  Note
that in the absence of GS~2023+34, a much narrower range of masses is
required.}

\end{figure}

\begin{figure}
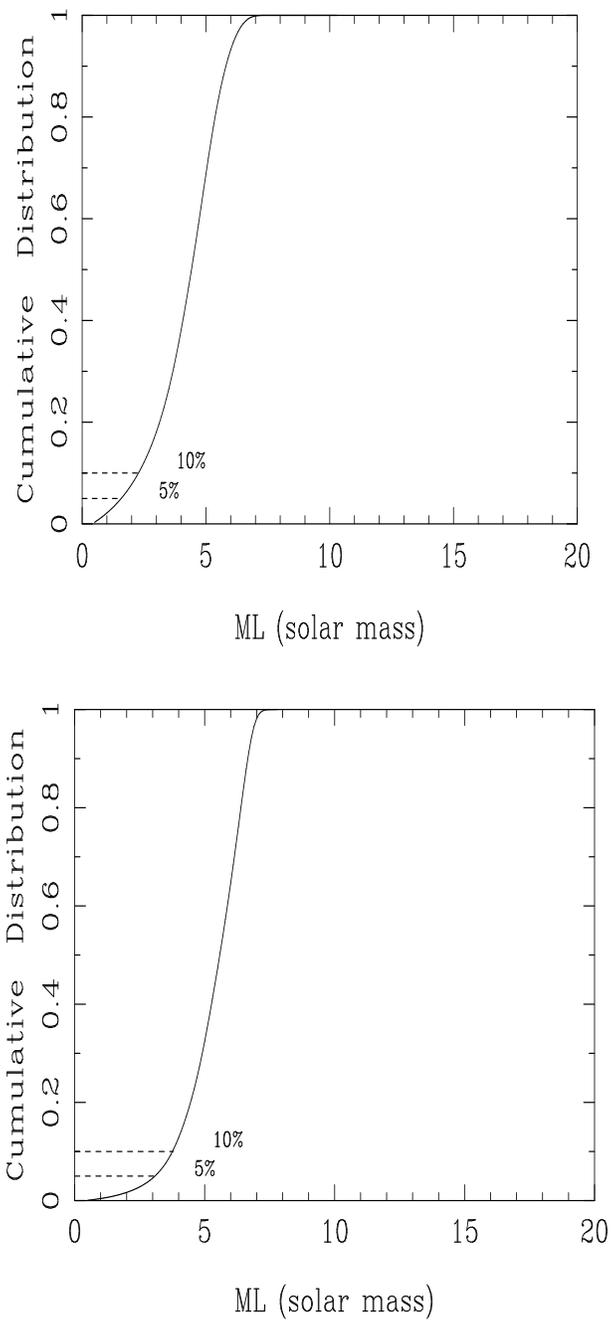

\centerline{
\psfig{figure=bc_all_3_ml.ps,height=9cm,width=8cm,angle=270}
}
\centerline{
\psfig{figure=bc_no2023_3_ml.ps,height=9cm,width=8cm,angle=270}
}

\caption[]{Cumulative probability that $m_L$ is below a given value. Top panel
includes GS~2023+34; bottom panel does not.  Note that in the absence of
GS~2023+34, there is a $>95\%$ probability that $m_L>3M_{\odot }$.
}

\end{figure}

\end{document}